\documentclass{webofc}
\usepackage[varg]{txfonts}   
\usepackage{extarrows}
\usepackage[colorlinks=true,linkcolor=black]{hyperref}

\newcommand{\nn}{\nonumber}

\newcommand{\vx}{{\mathbf{x}}}

\newcommand{\la}{\langle}
\newcommand{\ra}{\rangle}

\newcommand{\ve}{\varepsilon}
\newcommand{\vep}{\varepsilon}
\begin{document}
\title{Methods on compositeness and related aspects }
%
%

\author{\firstname{Jos\'e Antonio} \lastname{Oller}\inst{1}\fnsep\thanks{\email{oller@um.es}} 
}

\institute{Departamento de F\'{\i}sica, Universidad de Murcia, 30071 Murcia, Spain}

\abstract{
In many physical applications, bound states and/or resonances are observed, which raises the question whether these states are elementary or composite. Here we elaborate on several methods for calculating   the compositeness $X$ of bound states and resonances in Quantum Mechanics, and in  Quantum Field Theory by introducing particle number operators. For resonances $X$ is typically complex and we discuss how to get meaningful results by using certain phase transformations in the $S$ matrix. }
\maketitle
%
%

\section{Introduction}
\label{221019.1}

We start by reviewing a few basic aspects of the problem of compositeness in hadron physics \cite{Weinberg:1962hj,Weinberg:1963zza,Weinberg:1965zz}. The Hamiltonian $H$ is first split in the free $H_0$ and interaction $V$ parts,
\begin{align}
  \label{221019.1}
  H=H_0+V~.
\end{align}
Both $H$ and $H_0$ share the same spectrum. For $H_0$, one has the bare eigenstates,
\begin{align}
  \label{221019.2}
 H_0|\varphi_\alpha\rangle&=E_\alpha|\varphi_\alpha\rangle~,~\text{ continuum spectrum}~,\nn\\
    H_0|\phi_n\rangle&=E_{B_n}|\phi_n\rangle~,~\text{ discrete spectrum}~,
\end{align}
where $|\phi_n\rangle$ is a bare ``elementary'' state,  and  $|\varphi_\alpha\rangle $ is made up by the direct product of free particles 
   (Greek letters are used as subscripts  to refer to  quantum numbers, among which one has the momentum, spin, etc). The physical spectrum is comprised by the eigenstates of  $H$,
  \begin{align}
  \label{221019.3}
    H|\psi^{\pm}_\alpha\rangle&=E_\alpha|\psi^{\pm}_\alpha\rangle~,~\text{ continuum spectrum}~,\nn\\
    H|\psi_{B_n}\rangle&=E_{B_n}|\psi_{B_n}\rangle~,~\text{ discrete spectrum}~.
  \end{align}
  Here, the $|\psi_{B_n}\ra$ are the bound states of the theory and the $|\psi^{\pm}_\alpha\ra$ are the scattering in/out states, respectively. Within this framework the definitions of compositeness $X$ and elementariness $Z$ of a bound state $|\psi_B\ra$ are as follows \cite{Weinberg:1962hj}.
  Let us consider the linear decomposition of this state in the basis of eigenstates of $H_0$,  
  \begin{align}
  \label{221019.4}
    |\psi_B\ra&=\sum_n \la \phi_n|\psi_B\ra|\phi_n\ra+\int d\alpha \la \varphi_\alpha|\psi_B\ra|\varphi_\alpha\ra~.
    \end{align}
    Then, the Parseval identity implies
\begin{align}
  \label{221019.5}
    \langle \psi_B| \psi_B\rangle=1&=\underbrace{\sum_n |\langle \phi_n|\psi_B\rangle|^2}_{\equiv Z}+\underbrace{\int d\alpha
    |\langle \varphi_\alpha|\psi_B\rangle|^2}_{\equiv X}~,\\
    1&=Z+X~.\nn
    \end{align}

A new interpretation based on the use of the number operators in the interaction or Dirac picture was introduced in Ref.~\cite{Oller:2017alp}, where more details can be found.  The basic idea is 
to take two free particles of types  $A$ and $B$ ($H_0|AB_\gamma\rangle=E_\gamma|AB_\gamma\rangle$).  The standard creation and annihilation operators of the two free particles are 
 $a^\dagger_\alpha$, $ a_\alpha$, $b^\dagger_\beta$, $b_\beta$, respectively.
In terms of them, the number operator $N_D$ for the total  number of these free particles is
\begin{align}
 \label{221020.1}
 N_D=\int d\alpha a_\alpha^\dagger a_\alpha+\int d\beta  b_\beta^\dagger b_\beta=N_D^A+N_D^B~,
\end{align}
with $N_D^{A}$ and $N_D^{B}$  the number operators for each species of particles separately. Here, the subscript $D$ refers to the Dirac picture used.  
In nonrelativistic Quantum Field Theory (QFT) one can  express these operators using the fields $\psi_A$ and $\psi_B$ as  \cite{Thirring:book1}
{\small \begin{align}
 \label{221020.2}
  &N_D=\int d^3x \left[
      \psi_A^\dagger (x)\psi_A(x)+ \psi_B^\dagger(x)\psi_B(x)\right]~,~x=(t,\vx)~,
  \end{align}
  Let us notice that in the interaction picture $[H_0,N_D]=0$, so that $N_D$ is actually time independent, $N_D(t)=N_D(0)$.  In terms of $N_D$ one can give an alternative definition for the compositeness $X$ as \cite{Oller:2017alp} 
    \begin{align}
 \label{221020.3}
 X=\frac{1}{2}\langle \psi_B|N_D|\psi_B\rangle~.
     \end{align}
The factor 1/2 is introduced because we are considering that only two-particle states made out of $A$ and $B$ may contribute to  $|\psi_B\ra$. Let us now probe the equivalence between the new definition of $X$ and the previous one in  Eq.~\eqref{221019.5}. From the linear decompositions of the bound-state $|\psi_B\ra$ in states $|AB_\gamma\ra$ and $|\phi_n\ra$ (eigenstates of $H_0$) we have 
   \begin{align}
 \label{221020.4}
     |\psi_B\rangle&=\int d\gamma C_\gamma |AB_\gamma\rangle+\sum_n C_n|\phi_n\rangle~,\\
    X&= \frac{1}{2}\langle \psi_B|N_D^A+N_D^B|\psi_B\rangle=\int d\gamma |C_\gamma|^2~.\nn
   \end{align}
   This definition is specially suitable for Effective Field  Theories (EFTs), like e.g. ChPT and hadron physics in general. 
   The point is that in a low-energy EFT written in terms of the (pseudo-)Goldstone bosons as the only degrees of freedom, it is valuable to define $X$ as in Eq.~\eqref{221020.3}, since it does not require  to have explicit  bare ``elementary'' states (fields) in the theoretical set up.

 This new definition is also the most adequate for the treatment in QFT since, as deduced in \cite{Oller:2017alp}, it can be rewritten as
\begin{align}
 \label{221020.5}
X=\frac{1}{2}\lim_{T\to +\infty}\frac{1}{T}\int d^4x 
\langle \varphi_B| P\left[ e^{-i\int_{-\infty}^{+\infty}dt' V_D(t')}
 \sum_i \psi_{A_i}^\dagger(x)\psi_{A_i}(x)
 \right] |\varphi_B\rangle~,
\end{align}
by expressing the number operators in terms of the nonrelativistic fields $\psi_{A_i}$. Compared with Eq.~\eqref{221020.4} here $A\equiv A_1$ and $B\equiv A_2$. 
Making use of the LSZ formalism one can directly express $X$ in terms of $S$-matrix elements, 
\begin{align}
 \label{221020.6a}
    X&\!=\frac{1}{2}\lim_{E\to E_B}\frac{(E-E_B)^2}{g_\alpha(k_B)^2} \lim_{T\to +\infty}\!\frac{1}{T}\int\! d^4x 
\langle \varphi_\alpha| P\left[ e^{-i\int_{-\infty}^{+\infty}dt' V_D(t')}
 \sum_i \psi_{A_i}^\dagger(x)\psi_{A_i}(x)
 \right] |\varphi_\alpha\rangle~,
  \end{align}
with $g_\alpha(k_B)$ the coupling of the bound state to the continuum in/out states. Of course, the previous equation clearly shows that $X$ is an observable in nonrelativistic QFT since it is expressed in terms of $S$-matrix elements in the presence of a local source term, made up by the number-operator density. 

\section{Explicit formulas}
\label{sec.221024.exp}

\begin{figure}
\centering
\sidecaption
 \includegraphics[angle=0.0,width=.6\textwidth]{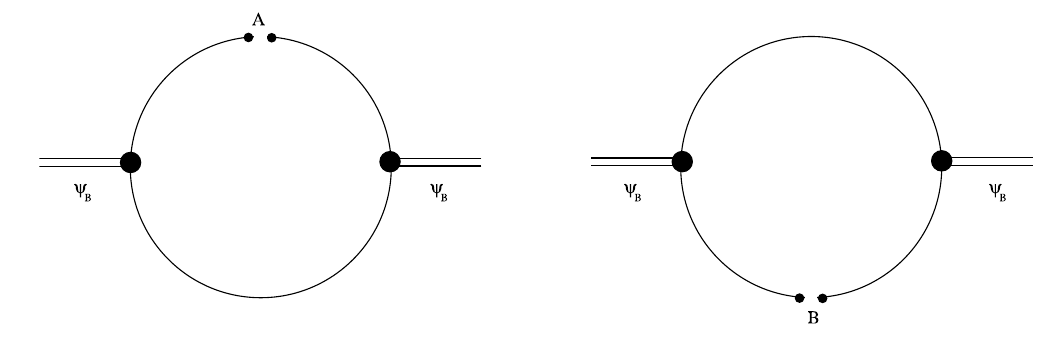}
 \caption{\small Feynman diagrams for the calculation of $X$ for a bound state represented by the double lines. The two filled dots correspond to insertions of
 a number-operator density.}
\label{fig.221020.1}       
\end{figure}

The Feynman diagrams appropriate for the calculation of Eq.~\eqref{221020.5} are depicted  in Fig.~\ref{fig.221020.1}, where the insertion of the number-operator density in the propagator of every particle is represented by the double dots. The resulting expression for $X$ in terms of the off-shell coupling  to  $A$ and $B$ is 
 \begin{align}
 \label{221020.6}
   X_{\ell S}&=\int\frac{d^3k}{(2\pi)^3}\frac{g_{\ell S}^2(k)}{(k^2/2\mu-E_B)^2}~,\\
   X&=\sum_{\ell S}X_{\ell S}~,\nn
 \end{align}
 with $\mu$ the reduced mass. The subscripts $\ell$ and $S$ in Eq.~\eqref{221020.6} refer to particular two-particle partial wave amplitude (PWAs)  coupled to the bound state (e.g. $^3S_1$ and $^3D_1$ for the deuteron in nucleon-nucleon ($NN$) scattering).
 If in the integral of Eq.~\eqref{221020.6} we assume a shallow bound state then the coupling $g_{\ell S}(k)$ can be factorized out with $k=0$, or with its on-shell value (which is valid  at this level of accuracy). Thus, $X_{\ell S}$ becomes
 \begin{align}
   \label{221021.9}
   X_{\ell S}&=g_{\ell S}^2\frac{d}{dE}\int\frac{d^3k}{(2\pi)^3}\frac{1}{k^2/2\mu-E_B}=g_{\ell S}^2\frac{i\mu^2}{2\pi k_B}~,
 \end{align}
 with the binding momentum $k_B=\sqrt{2\mu E_B}$. This is one of the results of Ref.~\cite{Weinberg:1965zz}. The equation for $g_{\ell S}(k)$ is obtained from the Lippmann-Schwinger equation $T=V+VGT$ in the limit $E\to E_B$ and by taking the residue at the pole, so that 
 \begin{align}
 \label{221020.7}
    g_{\ell S}(k)&
   =\frac{1}{2\pi^2}
   \int_0^\infty {k'}^2dk'V_{\ell S}(k,k')\frac{1}{{k'}^2/2\mu-E_B}g_{\ell S}({k'})~.
   \end{align}

 For the case of energy-independent potentials, which corresponds to  {\it pure potential scattering}, it was shown in Ref.~\cite{Oller:2017alp} that the compositeness is exactly equal to 1 for any bound state, that is,
 \begin{align}
 \label{221020.8}
\sum_{\alpha=1}^n X_\alpha=1~,
 \end{align}
 with $n$ the number of coupled PWAs. 
 We refer to \cite{Oller:2017alp} for the interesting detailed analysis, which is skipped here to avoid becoming too technical. 
 This result has interesting implications like e.g. in the case of the deuteron, studied within ChPT up to N$^4$LO in terms of energy-independent potentials with an estimated error in the reproduction of its properties of around a 4\% \cite{RodriguezEntem:2020jgp,Epelbaum:2014sza}. Therefore, one would conclude from Eq.~\eqref{221020.8} that for this bound state $X=0.96-1.0$.

  We now discuss within similar terms the case of resonances following the basic lines of scattering theory and the use of number operators  \cite{Oller:2017alp}. A resonance stems from the analytic continuation in energy of in states with energy $E+i\vep$, and out states  with energy $E-i\ve$ ($\ve\to 0^+$). In this way, when evaluating an $S$-matrix element $\langle \psi_\beta^-|\psi_\alpha^+\ra$ both  ket and bra  involve an energy of $E+i\vep$. In performing the analytical continuation towards the pole position of the resonance in the complex $E$-plane one has to cross the real axis along $E>0$, borrowing in the second Riemann sheet (RS) where the resonance pole lies with $\Im E<0$. 
For the calculation of $X$ for a resonance in  QFT we can then proceed similarly as for the case of a bound state by isolating the double pole residue  of  $S$-matrix elements with the external source $\sum_i \psi_{A_i}^\dagger(x)\psi_{A_i}(x)$. The corresponding formula is
\begin{align}
\label{221021.1}
    X=\frac{1}{2}\lim_{E\to E_R}\frac{(E-E_R)^2}{g_\alpha(k_R)^2}  \lim_{T\to +\infty}\!\frac{1}{T}\int\! d^4x 
\langle \varphi_\alpha| P\left[ e^{-i\int_{-\infty}^{+\infty}dt' V_D(t')}
 \sum_i \psi_{A_i}^\dagger(x)\psi_{A_i}(x)
 \right] |\varphi_\alpha\rangle~,
\end{align}
with $E_R$ and $k_R$ the resonance pole position in energy and momentum, respectively.   Of course, one can also write down a formula similar to Eq.~\eqref{221020.6} for a bound state, corresponding to analogous Feynman diagrams as those in Fig.~\ref{fig.221020.1} with the replacement $|\psi_B\ra\to |\psi_R\ra$. The result is
 \begin{align}
  \label{221021.3}
   X_{\ell S}&=\int\frac{d^3k}{(2\pi)^3}\frac{g_{\ell S}^2(k)}{(k^2/2\mu-E_n^2)^2}
   +\frac{i\mu^2}{\pi k_n}\left[\frac{\partial}{\partial k} k \,g_{\ell S}^2(k)\right]_{k=k_n}~,\\
   X&=\sum_{\ell S}X_{\ell S}~.\nn
 \end{align}
The extrapolation of the integral to the second RS is responsible for the additional last term as compared with Eq.~\eqref{221020.6}. 

In pure potential scattering it was proved in Ref.~\cite{Oller:2017alp} that $X=\sum_{\alpha=1}^n X_\alpha=1$, see also Ref.~\cite{Hernandez:1984zzb}. 
For instance, this would imply that $X$ for the virtual or antibound state in the $^1S_0$ $NN$ scattering should be very close to 1, since $NN$ scattering data can be described very precisely with energy-independent potentials derived from ChPT \cite{RodriguezEntem:2020jgp,Epelbaum:2014sza}. If for the deuteron, with a binding momentum of around 45~MeV, the possible uncertainty was bounded to a 4\%, for the virtual state in the $^1S_0$ PWA one would expect it to be smaller because $|k_R|\approx 10$~MeV$\ll 45$~MeV.


\section{Compositeness in the Heisenberg picture}

Let us denote by  $N_H(t)$ the number operator in the Heisenberg picture. The scattering states $|\psi^\pm\ra$ are eigenstates of $H$ that behave as free states when acted with operators at $\tau\mp\infty$, respectively \cite{Weinberg:book1}. Then,
\begin{align}
  \label{221025.1}
  \exp(-iH\tau)N_H(\tau)|\psi_\alpha^\pm\ra\to \exp(-iH_0\tau)N_D(\tau)|\varphi_\alpha\ra~,
\end{align}
for $\tau\to\mp\infty$. Taking into account in the previous equation that $|\psi^\pm\ra=U_D(0,\tau)|\varphi_\alpha\ra=\exp(iH\tau)\exp(-iH_0\tau)|\varphi_\alpha\ra$, with $\tau \to\mp\infty$, we have the equality at the operational level
\begin{align}
  \label{221025.1}
\lim_{\tau\to\pm}  \exp(-iH\tau)N_H(\tau)\exp(iH\tau)=N_D(\tau)~.
\end{align}
Since $H|\psi_\alpha^\pm\ra=E_\alpha|\psi_\alpha^\pm\ra$ we also have that 
\begin{align}
  \label{221025.2}
  \la \psi_\alpha^-|N_D(\tau)|\psi_\alpha^+\ra=
\la \psi_\alpha^-|N_H(\tau)|\psi_\alpha^+\ra~,~\tau\to\pm \infty.
\end{align}
Furthermore, as $d N_H(t)/dt=i[H,N_H(t)]$ it follows that
\begin{align}
  \label{221025.4}
\frac{d}{dt}\la \psi_\alpha^-|N_H(t)|\psi_\alpha^+\ra=0~,
\end{align}
so that both expectation values in Eq.~\eqref{221025.2} are time independent, and we can simply write that
\begin{align}  \label{221025.3}
\la \psi_\alpha^-|N_D(t)|\psi_\alpha^+\ra=
\la \psi_\alpha^-|N_H(t')|\psi_\alpha^+\ra~,
\end{align}
for arbitrary $t$ and $t'$. 
 The equality between the expectation values of $N_D$ and $N_H$ is also clear by taking into account the  QFT expressions above for $X$ in Eqs.~\eqref{221020.6a} and \eqref{221021.1}, since they  correspond to the $S$-matrix elements  $\displaystyle{\lim_{T\to\infty}\frac{1}{T}\la \psi_\alpha^-|\int_{-T/2}^{T/2} dt \,N_H(t)|\psi_\alpha^+\ra}$ when calculated in the interaction picture. Afterwards, the residue of the double pole is isolated. 

 Now, the idea would be to use an appropriate operator normalized in lattice QCD to mimic the number operators.  For instance, if the operator in the Heisenberg picture is $O(t)$ and we consider a bound state made out of nucleons, like the deuteron,  we could take the normalization constant to be $\langle N|O(t)|N \rangle$, which indeed is time independent (to conclude this one can follow an analogous reasoning as used for Eq.~\eqref{221025.4}).
 Of course, extra  discussions about the final adequate choice for the operator $O(t)$ in lattice QCD are still required, and this should be the object of further work. 



\section{Sum rule}

From a generic formula for two-particle  PWAs one can deduce a sum rule for compositeness and elementariness. We follow here Ref.~\cite{Guo:2015daa}, which employs relativistic kinematics.  
 Two-body unitarity along the right-hand cut, and above the thresholds of channels $i$ and $j$, can be expressed as \cite{Oller:2019rej}
\begin{align}
  \label{221021.4}
  \Im \left.T^{-1}\right|_{ij}=\delta_{ij}\rho_i~,
\end{align}
where $\rho_i$ is the phase-space factor $\rho_i=p_i/8\pi\sqrt{s}$, with $p_i$ the center of mass (CM) momentum for the channel $i$ and $s$ the CM total energy squared (the usual Mandelstam variable). From this equation one can resum the right-hand or unitarity cut and obtain a general  expression for a PWA in coupled channels. In matrix notation it reads \cite{Oller:2019opk,Oller:2019rej}
\begin{align}
  \label{221021.5}
  T(s)=\left[{\cal K}(s)^{-1}+G(s)\right]^{-1}~,
\end{align}
where ${\cal K}$ is an interacting kernel and the unitary loop function $G_i(s)$  is given by 
\begin{align}
  \label{221021.6}
G_i(s)&=\frac{1}{16\pi^2}\left[
a_i(\Lambda)+\log\frac{m_{i1}^2}{\Lambda^2}-x_+\log\frac{x_+-1}{x_+}-x_-\log\frac{x_--1}{x_-}\right]~,\\
x_\pm&=\frac{s+m_{i2}^2-m_{i1}^2}{2s}\pm\frac{1}{2s}\sqrt{(s+m_{i2}^2-m_{i1}^2)^2-4s(m_{i2}^2-i0^+)}~,\nn
\end{align}
where $m_{1i}$ and $m_{2i}$ are the masses of the two particles in the channel $i$, and $a_i(\Lambda)$ is a subtraction constant (the combination $a_i(\Lambda)-\log\Lambda^2$ is $\Lambda$ independent). 
Taking the derivative of Eq.~\eqref{221021.5} with respect to $s$, in the limit $s\to s_R$ the residue of the double pole implies that
\begin{align}
  \label{221021.7}
  1&=\underbrace{-\sum_i g_i^2 \left.\frac{dG_i(s)}{ds}\right|_{s_B}}_{X_i}+\underbrace{g^T G(s_B) 
  \left.\frac{dK(s)}{ds}\right|_{s_B}G(s_B) g}_{Z}~.
  \end{align}
The expression for $X_i$ does not depend on $a_i$ or  $\Lambda$, since they disappear in the derivative of $G_i(s)$. It is the same expression as in Eq.~\eqref{221021.9} for shallow bound states, and also for separable potentials \cite{Sekihara:2014kya,Aceti:2012dd}.

\section{Resonances}
\label{sec.221025.res}

For the case of resonances one should consider the unitarity loop function in the 2nd RS for those channels in which the transit to this sheet is involved to find the pole position at $s_R$. We indicate this by $G^{II}(s_R)$. Then, Eq.~\eqref{221021.7} implies
\begin{align}
  \label{221022.1}
  X_i&=- g_i^2 \left.\frac{dG_i^{II}(s)}{ds}\right|_{s_R}~.
  \end{align} 
However, for resonances $X_i$ is typically complex.
 We argue next that one should actually take its absolute value  $|X_i|$.
This based on certain phase transformations of the $S$-matrix in PWAs, driving to a phase redefinition of the couplings \cite{Guo:2015daa}.
 Let us first  consider a narrow resonance  $\Gamma\ll M_R-m_{\rm th}$, with $\Gamma$, $M_R$ the resonance mass and width, and $m_{\rm th}$ the nearest threshold. Later on we generalize these results for wider resonances. The $S$ matrix is split in a pole term plus $S_0(s)$, which is assumed to be smooth in a certain domain around the resonance pole. That is,
\begin{align}
  \label{221021.8}
  S(s)=\frac{R}{s-s_R}+S_0(s)~,
\end{align}
with $R$ the residue matrix of the resonance pole at $s_R$.  
Then, we impose unitarity of the $S$ matrix in PWAs,
\begin{align}
S(s)S(s)^\dagger=I~.
\end{align}
The solution of this equation with $S(s)$ given by  Eq.~\eqref{221021.8}, with constant $S_0$ for $s$ near $s_R$, is 
\begin{align}
  \label{221021.10}
  S(s)&={\cal O}\underbrace{\left(
I + \frac{ i \lambda {\cal A} }{s-s_R} 
\right)}_{S_R(s)}{\cal O}^T~,\\
  {\cal O}{\cal O}^\dagger&=I~.\nn
\end{align}
Here ${\cal A}$ is a rank one symmetric projection operator and
$S_R(s)$ is the pure resonant $S$ matrix. The disposition of operators in Eq.~\eqref{221021.10} clearly indicates the  corrections to  $S_R$ due to initial- and final-state interactions from $S_0$.


  \begin{figure}
\centering
\sidecaption
 \includegraphics[angle=0.0,width=.4\textwidth]{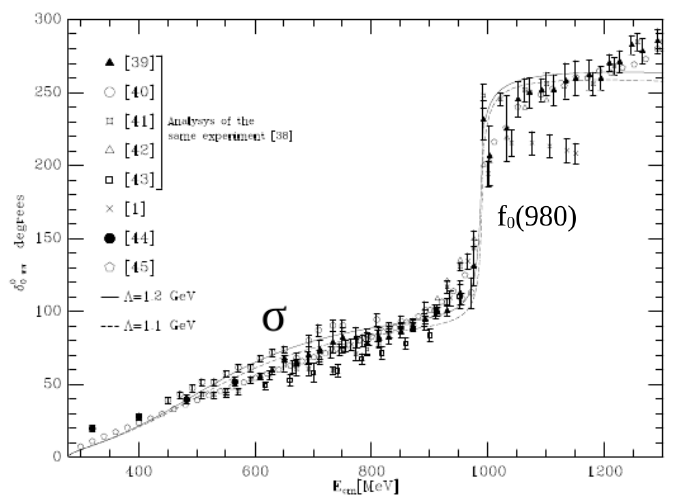}
 \caption{\small Isoscalar scalar $\pi\pi$ phase shifts, $J^{PC}=0^{++}$. For the experimental references see \cite{Oller:1997ti}.}
\label{fig.221021.3}       
  \end{figure}

 The dressing to $S_R$ by $S_0$ typically modifies the phases of the resonance couplings in $S_R$. This is because the moduli of the couplings $g_i$ to the different channels have physical meaning, since they provide the partial decay widths of a narrow resonance \cite{ParticleDataGroup:2022pth}, $\Gamma_i={|g_i|^2}/({8\pi M_R^2})$.  Then, the $S$-matrix phase transformation only change the phases of the resonance couplings
  \begin{align}
  \label{221021.13}
   S_{\cal O}(s)&\equiv {\cal O} S(s) {\cal O}^T~,\\
   {\cal O}&={\rm diag}(e^{i\phi_1},\ldots,e^{i\phi_n})~,\nn \\
   g_i^2&\to g_i^2 e^{2i\phi_i}~.\nn
  \end{align}
As an example of this mechanism let us consider $\pi\pi-K\bar{K}$ scattering in $J=0$ and $I=0$ ($I$ is the total isospin). As the author checked in the calculations for Ref.~\cite{Oller:1997ti}, the coupling to $\pi\pi$ of the $f_0(980)$ is a positive number times a phase-factor $i$, which stems from the phase-shifts of $\pi\pi$ rescattering due to the $f_0(500)$ resonance  at the rise of the $f_0(980)$. This is clearly seen from Fig.~\ref{fig.221021.3}. 

Therefore, $X_i$, Eq.~\eqref{221022.1},  only changes its phase under the phase transformations of Eq.~\eqref{221021.13}, and they can be chosen so that it becomes a real positive number. Hence,  
\begin{align}
  \label{221021.14}
X_i \to & |X_i|\geq 0~.
\end{align}
Let us now generalize Eq.~\eqref{221021.14} for non-narrow resonances. The point is to require the validity of the Laurent series expansion around the resonance pole up to physical values of $s$ above  threshold, so that the resonance couplings still directly impact physical energies. This implies that $s_{{\rm th};n}<\Re s_R<s_{{\rm th};n+1}$, where $s_{{\rm th};n}$ is  the threshold of channel $n$. 

\section{Effective range expansion}
\label{sec.221023.ere}
We assume now that a bound or resonance pole lies close enough to a relevant threshold so that the well-known effective range expansion (ERE) can be applied. We follow here the developments in Ref.~\cite{Kang:2016ezb}. Up to and including the scattering length $a$ and the effective range $r$ the $S$-wave PWA can be written as
  \begin{align}
  \label{221021.15}
    T(k)&=\frac{1}{-\frac{1}{a}+\frac{1}{2}r k^2+G(k)}~,~~~
    G(k)=-i k~.
  \end{align}
  Then, if $T(k)$ has a resonance pole at $E=E_R=M_R-i\Gamma/2$ (energy is measured from $m_{\rm th}$),  $a$ and $r$ can be given in terms of the mass and width of the resonance as
\begin{align}
  \label{221021.16}
a&=-\frac{2 k_i}{|k_R|^2}~~,~~k_R=\sqrt{2\mu E_R}=k_r-i\,k_i~,\\
r&=-\frac{1}{k_i}~~,~~\text{so that }\frac{r}{a}>2~.\nn
     \end{align}
By applying Eq.~\eqref{221022.1} we can express the compositeness as
\begin{align}
  \label{221022.2}
X&=-\gamma^2\frac{dG}{ds}=-\gamma_k^2\frac{dG}{dk}=i\frac{k_i}{k_r}=i\tan\frac{\phi}{2}~,
\end{align}
with $\gamma_k^2$ the residue at the pole position in the complex $k$-plane.
The simplicity of the ERE up to including $r$  perfectly illustrates the condition on the mass of the resonance, discussed at the end of Sec.~\ref{sec.221025.res}, because 
  \begin{align}
  \label{221022.3}
  |X|\leq 1 \leftrightarrow \phi\in [0,\pi/2]  \leftrightarrow
  M_R\geq 0~,
  \end{align}
  with $|X|=1$ for $M_R=0$ and $\Gamma>0$.
  Furthermore, since $X$ is purely imaginary in Eq.~\eqref{221022.2}  if its real part is taken to end with a real compositeness, as advocated in Ref.~\cite{Aceti:2014ala}, then from the ERE $X=0$. Of course, this result is not meaningful  in such generality. The expression for $|X|$ can also be rewritten as $|X|=\left(\frac{2r}{a}-1\right)^{-1}$. In the subsequent we drop the sign of modulus on $X$ and directly consider by this symbol  the compositeness calculated within the ERE.

\section{CDD poles. Track of elementariness}
\label{sec.221022.cdd}

While the ERE is a smooth expansion  around $k=0$ of the combination $t(E)^{-1}+ik$, which has no right-hand cut, more structure can be accounted for by allowing poles in $t(E)^{-1}$. These are  the so-called CDD poles \cite{Castillejo:1955ed}.  
 In this way, a once-subtracted dispersion relation (DR) for $t(E)^{-1}$ accounting for a pole in $t(E)^{-1}$  at $M_Z$ gives 
\begin{align}
  \label{221022.7}
  t(E)=\frac{1}{\frac{\lambda}{E-M_Z}+\beta-ik}~,
\end{align}
with $\lambda$ the residue at the CDD pole and $\beta$ a subtraction constant.
A consequence of this pole is that the ERE or a Flatt\'e parameterization break down for $|k|\geqslant \sqrt{2\mu |M_Z|}$. 
A near-threshold CDD pole contributes to   $a$ and $r$ as
  \begin{align}
  \label{221022.8}
    \delta a&=\frac{M_Z}{\lambda}~,~~~  \delta r =-\frac{\lambda}{m M_Z^2}~.
  \end{align}
  Therefore, $\delta a \to 0$ and $\delta r\to -\infty$ (unless $\lambda=0$) for $M_Z\to 0$. As a result,
  a large negative value of $r$ is a hint for a nearby CDD pole and for elementariness. Indeed, if we calculate $X$ from the ERE formula $X=\left(\frac{2r}{a}-1\right)^{-1}\to 0$ for $M_Z\to 0$. 
  Nonetheless, one should keep in mind that having a near CDD pole to a resonance pole is sufficient but not necessary for the resonance to be elementary. For instance, as analyzed in Ref.~\cite{Oller:1998zr}, the CDD pole associated to the $\rho(770)$ resonance 
 is located at $s=M_\rho/(1-g_V^2)$ and tends to infinity because of the KSFR relation, which requires $g_V^2=1$. This is so despite the resonance $\rho(770)$ is a clear example of a $q\bar{q}$ resonance. From a hadronic point of view this is clearly signaled by studying the dependence of the pole position with the number of colors of QCD $N_c$ \cite{Pelaez:2006nj,Guo:2012yt}.

\section{Determination of $X$ by making use of the decays of the resonance}
\label{sec.221022.1}
Here we outline the method introduced in Ref.~\cite{Meissner:2015mza}
to combine the knowledge of the resonance pole position and the saturation of its width and branching ratios (when they are available), and/or of the compositeness $X$. This method has the advantage that it is a coupled-channel study, while the one based in the ERE takes into account only one channel. If the channel 1 is the lighter channel and 2 is the one near the mass of the resonance, the main equations are $X=X_1+X_2$ and $\Gamma=\Gamma_1+\Gamma_2$, with 
\begin{align}
\label{221022.4}
\Gamma_1&=\frac{2 X_1 }{\mu} k(M_R)|k_R|~,~~~\Gamma_2=\frac{X_2|k_R| M_R^2}{\pi \mu}
\int_{M_{\rm th}}^{+\infty}dW\frac{k(W)}{W^2}\frac{\Gamma}{(M_R-W)^2+\Gamma^2/4}~.
\end{align}
From these equations one can then constrain the couplings or, equivalently, the  partial compositeness coefficients.

As a concluding remark, we  stress that the different methods  used in many instances to study the nature of resonances have provided when used simultaneously a consistent picture on the nature of these resonances. Several examples are discussed in Ref.~\cite{oller.theseproceedings}. This is interesting since each method has its own realm of applicability, and the fact that one can achieve compatible conclusions from all of them reinforces the global picture that emerges. 


     \section*{Acknowledgements} 
This work has been supported in part by the MICINN
AEI (Spain) Grant No. PID2019–106080GB-C22/AEI/10.13039/501100011033\,. I thank Zhi-Hui Guo for the feedback on the manuscript.

\bibliography{references}
\end{document}